\documentstyle[floats,psfig,aps]{revtex}
\def\be{\begin{equation}}
\def\ee{\end{equation}}
\def\bea{\begin{eqnarray}}
\def\eea{\end{eqnarray}}

\begin{document}

%\draft
\twocolumn[

\title{The multifragmentation of spectator matter}

%%%%%%%%%%%%%%%%%%%%%%%%%%%%%%%%%%%%%%%%%%%%%%%%%

\author{P.B. Gossiaux, R. Puri, Ch. Hartnack, and J. Aichelin}

%%%%%%%%%%%%%%%%%%%%%%%%%%%%%%%%%%%%%%%%%%%%%%%%%

%%%%%%%%%%%%%%%%%%%%%%%%%%%%%%%%%%%%%%%%%%%%%%%%%

%%%%%%%%%%

\address{SUBATECH \\
Laboratoire de Physique Subatomique et des Technologies Associ\'ees \\
UMR Universit\'e de Nantes, IN2P3/CNRS, Ecole des Mines de Nantes\\
4, rue Alfred Kastler 
F-44070 Nantes Cedex 03, France.}

%%%%%%%%%%%%%%%%%%%%%%%%%%%%%%%%%%%%%%%%%%%%%%%%%

%%%%%%%%%

%

\maketitle

\widetext

\begin{quote}
\begin{abstract}
We present the first microscopic calculation of  the spectator fragmentation  
observed in heavy ion 
reactions at relativistic energies which reproduces the
slope of the kinetic energy spectra of the fragments as well as 
their multiplicity, both measured by the ALADIN collaboration.
In the past both have been explained in thermal models, however with 
vastly different assumptions
about the excitation energy and the density of the system.  We show that both observables are
dominated by  dynamical processes  and  that the system does not pass a state of  thermal
equilibrium. These findings question the recent conjecture that in these
collisions a phase transition of first order, similar to that between water 
and vapor, can be observed. 
\end{abstract}

\end{quote}

\narrowtext]

%\pagebreak

%%%%%%%%%%%%

%\psdraft

%%%%%%%%%%%%%%%%%%%%%%%%%%%% Introduction%%%%%%%%%%%%%%%%%%%%%%%%

\section{Introduction}
In semiperipheral reactions of relativistic heavy ions  it seems that geometry determines the scenario.
The overlapping nucleons of projectile and target form a fireball 
of high temperature and a density
well above normal nuclear matter density. The non overlapping nucleons, called spectators, gain 
little excitation energy and  continue their way with almost the original velocity.  This is a conclusion
one has drawn from many experiments, first in Berkeley  and later at GSI.  Whereas the physics of the 
fireball
is pretty well understood that of the spectators is still under debate. Up to now, the most
complete experiments of the ALADIN collaboration have shown that the spectator matter disintegrates
into many intermediate mass fragments (IMF's). Above 400 MeV/N 
the fragment multiplicity distribution  
for a given projectile-target combination does not change anymore. For a 
for a given projectile the 
average number of  projectile like IMF's as
a function of the  total bound charge  becomes  independent 
of the target\cite{schu}. This dependence as well as the fragment multiplicity
distribution have been almost perfectly described by  a thermodynamical model, 
advanced by Botvina et 
al. \cite{bot}, under the assumption that 
the excitation energy of the spectator matter is of the 
order of a couple of MeV/N.

An independent method to measure the excitation energy (or more exactly
the temperature) has been employed by the ALADIN collaboration: 
measuring ratios of isotope yields and assuming that they are produced in thermal
equilibrium, one can determine the temperature, which turns out to be about 5 MeV,
almost independently of the ratios used. 
If one displays this temperature as a function of the excitation energy,
we observe a functional dependence typical for a first order
phase-transition. Consequently, the ALADIN collaboration conjectured
that in these reactions a transition from a liquid 
phase of nuclear matter to a gaseous phase takes place \cite{poc}.  

This temperature of about 5 MeV is in sharp contrast to the one extracted
from the slopes of the fragment velocity spectra, measured in the 
very same experiment. They indeed show a thermal shape but the observed slopes 
correspond to a temperature of 30 MeV, a value about 6 times larger than that 
extracted from the particle unstable states or the isotope ratios.
Thus it seems that several observables show a functional form as 
expected for a  system in thermal equilibrium but that the temperatures 
differ by large factors.

Up to now, dynamical models like the Quantum Molecular Dynamics approach (QMD) 
\cite{aic} have not succeeded in 
describing the data. They underestimate by far the multiplicity of 
IMF's \cite{hub} 
and have therefore not been considered as reliable despite their success to 
describe multifragmentation data
at low beam energies. The reason for this failure remained obscure 
because it has been verified that for a given excitation energy 
a single nucleus disintegrates in  
QMD simulations into the same number of fragments as in 
standard statistical models 
\cite{beg}.  Thus, one could conclude that in QMD simulations 
not sufficient energy is transferred to the spectator matter,
a puzzling result because the main mechanism for the energy transfer, 
the elastic NN collisions, are well under control and can be cross-checked
by comparison with experimental rapidity distributions. 

In QMD calculations each nucleon $\alpha$ moves on a classical trajectory 
as obtained by a variational solution of the n - body Schr\"odinger equation:  
\be 
\dot{\vec p_{\alpha}} = - \vec \nabla_{\vec x_a} \sum_\beta <V(
\vec x_\alpha, \vec x_\beta)>
\ee
and                  
\be 
\dot{\vec x_{\alpha}} =  \vec p_\alpha/m.
\ee
where $\vec p_{\alpha}$ and $\vec x_{\alpha}$ are the centroids of the
Gaussian wave functions (in momentum and coordinate space respectively)
which represent the nucleons. $V(
\vec x_\alpha, \vec x_\beta)$ is the two-body interaction between the
nucleons. In addition the nucleons interact via stochastic elastic and
inelastic NN collisions. For details of the
approach we refer to ref. \cite{aic}.

Recently we advanced a new algorithm \cite{pur} for fragment 
recognition which allows to identify fragments at any time during the
reaction. The up to then used Minimum Spanning Tree (MST) method
requires the fragments
to be well separated from each other in coordinate space,
limiting the fragment identification to a very late stage of the 
reaction only. MST assumes that two nucleons are part of a common fragment
if their distance $\sqrt{(\vec x_\alpha - \vec x_\beta)^2}$
is smaller than $r_C = 3 fm$. This new
algorithm is called Simulated Annealing Cluster Algorithm (SACA).
It searches for that configuration of nucleons and clusters with
the largest total binding energy $ \sum \zeta_k$, where $\zeta_k$
is defined as  
%%%%%%%%%%%%%%%%%%%%%%%%%%%
%
\begin{equation} \label{Ebind}
\zeta_k =  \frac{1}{N^f}\left[\sum_{i=1}^{N^f} 
\frac{(\vec p_i - \vec p_{cm})^
{2}}{2m}+
\frac{1}{2} \sum_{i\ne j}^{N^{f}} V_{ij}\right].
\end{equation}
%
%%%%%%%%%%%%%%%%%%%%%%%%%%%%%%%%%%%%%%%%
provided that
$\zeta_k <  -4.0$ MeV and  $N^f \ge$ 3 and as zero otherwise.  In
this definition, $N^f$ is the number of nucleons in a fragment and 
$\vec p_{cm}$ its center-of-mass momentum. For the technical
aspects of this algorithm we refer to \cite{pur}. This algorithm 
has overcome the mathematical difficulties to treat systems of the size of 
400 nucleons with a simulated annealing algorithm which limited the
similar approach advanced in \cite{dor}. It is also similar in spirit to the 
cluster
recognition algorithm advanced by Garcia et al. \cite{gar} which, however,
does not always find the most bound configuration but may be caught in a 
local minimum.
 
This new algorithm allows for new insights into the reaction mechanism,
especially into the time dependence of the fragment formation. In fig. 1
we display the time evolution of different observables 
for the reaction Au + Au 600 MeV/N, b = 8 fm, comparing MST with
SACA. This impact parameter has been chosen because it yields the
largest IMF multiplicity.

The first row shows the collision rate and the density averaged over the centroids of all nucleons
$$ \rho= {1\over N} \sum^N_{\alpha=1} \rho (\vec x_\alpha)= 
{1\over N}\sum^N_{\alpha=1}\sum^N_{\beta=1}({1 \over 2\pi L})^{3/2}
e^{-(\vec x_\alpha - \vec x_\beta)^2/2L}$$
with $x_\alpha$ being the centroid position of the nucleon $\alpha$ and L has the
standard value of $L = 1.08 fm^2$. Please note that in our definition of the density
a free particle still has a "self density" of $({1 \over 2\pi L})^{3/2}=
.32 \rho_0$. The second row shows the time evolution of the average mass of 
the heaviest fragment.
According to SACA, it first decreases to a minimum at 
t = 50 fm/c and re-increases later. Looking in details, one understands 
that this re-increase is mostly due to the reabsorption of some of the IMF's, as 
displayed in the third row. At the same time
the largest fragment looses nucleons due to evaporation. Hence 
at the end of the violent phase of the reaction SACA finds that 
the spectator matter is 
subdivided into small fragments. They are, however, located in the same space
region,
as can be seen from the large value of $A_{max}$ obtained by the MST procedure.
Nucleons of different fragments interact and proceed towards a thermal 
equilibrium. The fragments loose their individual identity and 
merge into a large cluster. Consequently, the size of $A_{max}$ increases.
 
The persistence coefficient is studied in the last row of fig.~1.
This concept has been introduced in \cite{pur} as follows:
We first define the number of pairs of nucleons in the cluster $C$ at time t 
$b_C(t)=N_C(N_C-1)/2$. At a time $\Delta t$ later, some of the nucleons 
may have left the cluster and are part of another cluster or singles
and others may have entered the cluster. 
Let $N_{C_A}$ be the number of nucleons which have been in the cluster 
C at time t and are at $t+\Delta t$ in the cluster A. 
We define $a_C(t+\Delta t)=\sum_A0.5*N_{C_A}(N_{C_A}-1)$, 
where the sum goes over all clusters A present at time $t+\Delta t$. 
The persistence coefficient of the cluster $C$ is now defined as 
\begin{equation}
$$P_C(t+\frac{\Delta t}2)= a_C(t+\Delta
t)/b_C(t)$$
\end{equation}

The mean persistence coefficient for the fragments $2 \le A \le 4$
and $4 < A \le 65$ is displayed in the last row of fig.~1.
Its low value as well as the considerable differences between results deduced with SACA 
and MST (performed at $t > 200 fm/c$) indicate that after 60 fm/c
the fragments are still very close in coordinate space and exchanging particles. 
We will come back to this later.

%%%%%%%%%%%%%%%%% Fig .1 %%%%%%%%%%%%%%%%%%%%%%%%%%%
\begin{figure}[htb]
\centerline{\psfig{figure=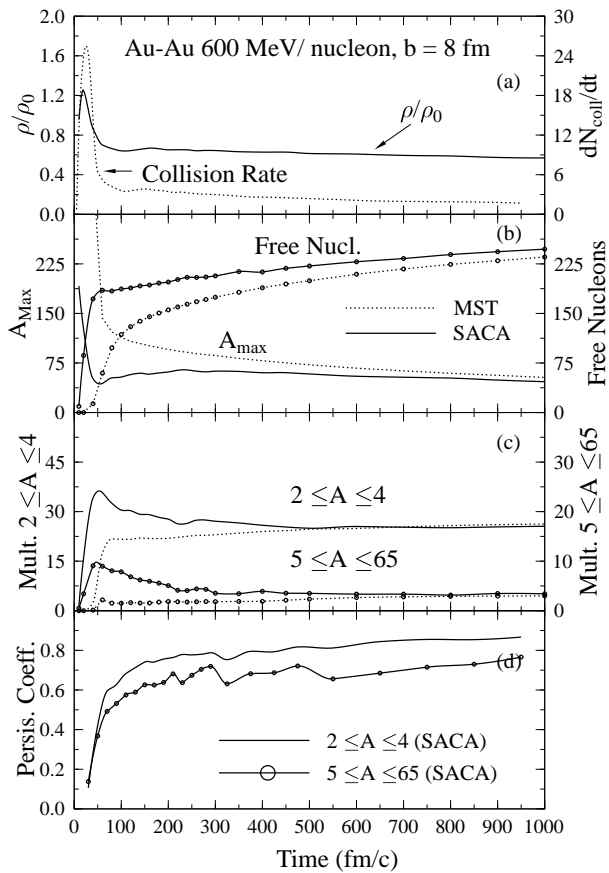,width=0.95\hsize}}
\caption{Evolution of the collision of Au-Au at 600 MeV/N and at
an impact parameter of b = 8 fm. From top to bottom the graphs display
the time evolution of the density and of the collision rate,
of the size of the heaviest fragments and of the number of 
emitted nucleons, of the multiplicity of fragments
with masses $ 2\le A \le 4$ and with masses $ 5 \le A \le 65$ 
and of the persistence coefficient.}
\end{figure}
%%%%%%%%%%%%%%%%% Fig .1 %%%%%%%%%%%%%%%%%%%%%%%%%%%
Due to the width L of the Gaussian wave function
the expectation value of the two-body potential has a range of 3.6 fm which is 
large as compared to the
range of the nuclear force in free space. One may therefore conjecture
that (1) the late thermalization of the spectator matter is due to this large 
range and therefore fictitious as well as its consequences (late decrease of the number of
IMF's,etc.) and (2) that in reality the spectator matter
breaks apart into the IMF's observed at t = 60 fm/c. In this
article we present results based on this key conjecture. 

In figure 2 we 
display the IMF multiplicity
distribution and the size of the largest fragment obtained after 60 fm/c.
Both are compared with
the experimental data as well as with the results obtained with the 
MST method at t=200 fm/c. As can be seen, 
the multiplicity distribution obtained with SACA
agrees quite well with the data whereas the MST multiplicity distribution
fails completely as already observed in \cite{hub}. 
%%%%%%%%%%%%%%%%% Fig .2 %%%%%%%%%%%%%%%%%%%%%%%%%%%
\begin{figure}[htb]
\centerline{\psfig{figure=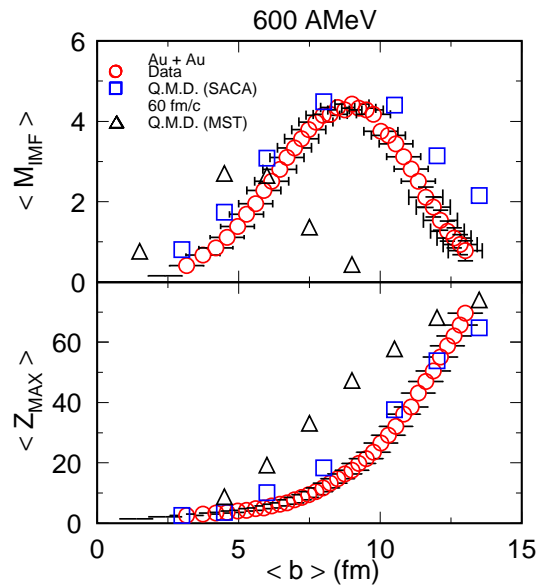,width=0.95\hsize}}
\caption{
Multiplicity of IMF's and the charge of the largest fragment
as a function of the impact parameter. We show the experimental data 
[10],
the QMD calculation using the minimum spanning tree methods and 
using SACA at the moment the size of the largest fragment reaches 
its minimal value, as explained in the text.}
\end{figure}
%%%%%%%%%%%%%%%%% Fig .2 %%%%%%%%%%%%%%%%%%%%%%%%%%%

We proceed now to the dynamical variables. The mean value and the
variance of the experimental rapidity distribution of the fragments \cite{sch}
as a function of the fragment mass   
are displayed in fig.3. The fragments have a rapidity close 
to the beam rapidity which is marked by a dashed line. 
Because the fragment energy spectra
look very much like the ones expected from a system in thermal equilibrium 
and because the variance of the rapidity
decreases with the mass number roughly like $\sigma(y) \propto {1\over 
\sqrt{A}}$
one is tempted to extract a "temperature"  $T = \sigma^2 (y)\cdot A \cdot m_N$.
The value one obtains is about 30 MeV. If one follows the conjecture
that multifragmentation is a thermal process one is confronted with the
fact that the gas of fragments of a given size A has a 
6-8 times higher "temperature" than the spectator matter at the moment
the fragments are formed. As only a small fraction of this difference
can be attributed to the Coulomb repulsion between the fragments, 
a yet unknown mechanism is required for heating up the 
gas of fragments after creation. Such a heating of the system
to a temperature of 30 MeV would require more energy than available,
as it can be calculated from the kinetic energy of the projectile spectator
matter. Furthermore, the high temperature
of the fragment kinetic energy spectra raises 
the question how fragments can survive at all in this hot environment and
do not disintegrate into nucleons.
 
In fig. 3 we display  the filtered results of the QMD calculation as well and
find that calculation and experiment agree quite well. This is the
first time that both the variance of the rapidity distribution and the 
multiplicity of IMF's have been reproduced by the same theoretical approach,
in contrast to the thermal models where both can be individually reproduced, 
however with vastly different temperatures.

Where does the differences of the apparent temperatures come from? 
We start our investigation with fig.4 where the time evolution of the
average rapidity and of the average transverse momentum in the reaction plane
of different classes of projectile-like fragments are displayed. 
"All nucleons" means that we average over
all nucleons which are entrained in fragments of the selected size 
(which is equivalent to the average over the rapidity or transverse
momentum of all fragments of the selected size). "Proj. nucleons" means
that we calculate the average value using exclusively the nucleons which have
been part of the projectile initially (which is by far the majority).    
This separation into projectile and target nucleons seems to be impossible as
soon as the nucleons have suffered their first collision and quantum mechanics
does not allow anymore to determine their origin. However is has been shown recently \cite{gos}
that, due to the forward-backward enhancement of the NN cross-section,
one arrives at the same conclusion if one always names that particle,
which is closest to the beam direction, "projectile" nucleon and vice versa.
By "Weighted", we mean that all fragments $A\ge 5$ are taken into account,
with their mass A as a weight. 
%%%%%%%%%%%%%%%%% Fig .3 %%%%%%%%%%%%%%%%%%%%%%%%%%%
\begin{figure}[htb]
\centerline{\psfig{figure=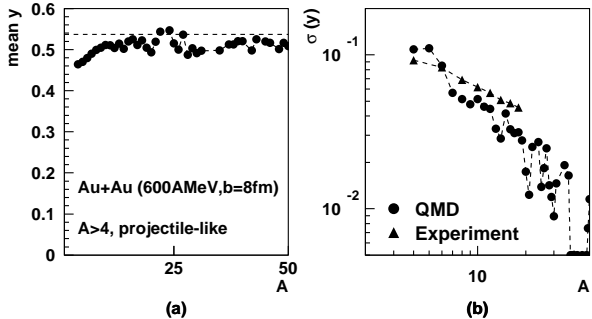,width=0.95\hsize}}
\caption{Mean rapidity and variance of the rapidity distribution
as a function of the fragment mass for Au-Au at 600 MeV/N.}
\end{figure}
%%%%%%%%%%%%%%%%% Fig .3 %%%%%%%%%%%%%%%%%%%%%%%%%%%
We observe that initially the projectile nucleons are decelerated 
substantially. Later the small IMF's are
accelerated again. This means that the nucleons which are entrained
in the projectile-like fragments have a substantial interaction with 
the target. If one takes the average over "all nucleons"
the deceleration and the acceleration of the IMF's balance each other for
the small fragments. The deceleration is caused by the interaction 
between projectile and target whereas the acceleration 
is caused by the interaction of the projectile-like fragments with the rest 
of the projectile-like spectator matter. The smaller the fragments the
more probable they are produced close to the projectile-target interface 
and the more they are decelerated (by potential
interaction or by absorbing a target nucleon). Later, the spectator matter,
which is less decelerated, re-accelerates these fragments.
In fact, the acceleration is more pronounced for small fragments which are 
produced close to the intersection and have suffered a larger deceleration
than the larger fragments. On the other hand, the small fragments
are mostly located in the tail and decouple pretty soon (80 fm/c) from the 
rest of the spectator matter: asymptotically, they remain slower. 
If one sums over all spectator fragments the
average rapidity has to be constant because forces between projectile-like 
fragments do not change the center-of-mass motion of the projectile-like spectator 
matter. This is indeed the case as can be seen in the bottom of figure~4.

The average transverse momentum remains very
small in the whole process. After the mutual attraction of projectile
and target we observe the well known bounce-off caused
by the higher density at the interface. As already discussed in ref.~\cite{gos} 
the large projectile-like fragments have always a surplus of 
nucleons which have a transverse momentum pointing away from the target, because those
having the opposite direction have a higher change to take part in the fireball. 
Hence a part of the observed final transverse momentum is due to an initial-time 
selection of the nucleons and not generated during the interaction. 
On the average, the light fragments are closer to the projectile-target overlap;
therefore, the bounce-off contribution is stronger. Besides, the initial-time contribution 
is less pronounced because these fragments may also absorb some target nucleons, which 
must have a strong transverse momentum pointing towards the projectile---in order to 
entrain it---.
%%%%%%%%%%%%%%%%% Fig .4 %%%%%%%%%%%%%%%%%%%%%%%%%%%
\begin{figure}[htb]
\centerline{\psfig{figure=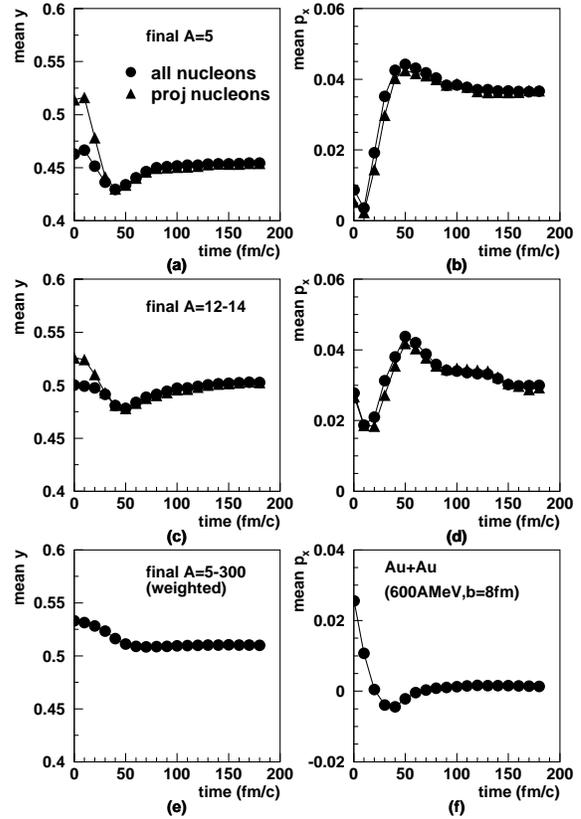,width=0.95\hsize}}
\caption{Evolution of the collision of Au-Au at 600 MeV/N at
an impact parameter of b = 8 fm. In the top and middle row we see the mean rapidity
and the mean $p_x$ momentum per nucleon of A=5 and A = 12-14 fragments, respectively. 
The last row display the average over all projectile like fragments.}
\end{figure}
%%%%%%%%%%%%%%%%% Fig .4 %%%%%%%%%%%%%%%%%%%%%%%%%%%
The variances of the momentum distributions of the fragments are displayed 
in fig.~5. They
are always calculated with respect to the mean values displayed in fig.~4.
Regarding "all nucleons" we see, first of all, that the variance 
for the small fragments as
well as that for all fragments changes little in the course of time. 
Hence, in contradistinction
to thermal models, this large variance is there from the beginning 
and not due to the randomization of collective motion in the course 
of the approach to thermal equilibrium.
The variance of the projectile nucleons changes, however,
considerably in the course of time. This is a consequence of the 
observation that the deceleration and hence the average
velocity of the projectile nucleons depends on their transverse 
distance to the participant region. 

If one separates three classes of A=5 fragments according to the
number of entrained target nucleons one finds
a mean rapidity close to $<y> ={ n_i\cdot y_{proj} + (5-n_i)\cdot y_{targ} 
\over 5}$
where $n_i$ is the number of projectile nucleons entrained in the
projectile like fragment. This can been seen in fig.~6 where we 
display the mean values and variances for A=5 projectile like fragments,
separately for 0, 1 or 2 entrained target nucleons. 

We observe that the entrained target nucleons
increase the transverse momentum of the fragment. It is obvious from geometry
that in order to get entrained in a projectile like fragment the target nucleons
must have a positive $p_x$ value. This momentum is added to that caused
by the bounce-off, which is around 40 MeV/c.

%%%%%%%%%%%%%%%%% Fig .5 %%%%%%%%%%%%%%%%%%%%%%%%%%%
\begin{figure}[htb]
\centerline{\psfig{figure=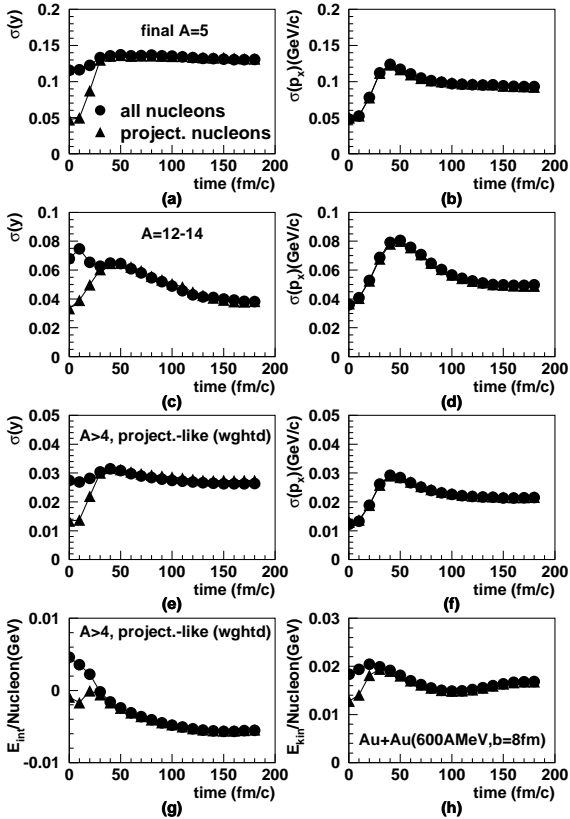,width=0.95\hsize}}
\caption{Evolution of the collision of Au-Au at 600 MeV/N at
an impact parameter of b = 8 fm. In the top and second row we see the 
variances of the rapidity and the $p_x$ momentum distributions
of A=5 and A = 12-14 fragments, respectively. 
The third row displays the average over all projectile like fragments.
The last row displays on the left hand side the average internal
energy of the fragments and on the right hand side the average kinetic
energy of the fragment nucleons in the fragment center of mass system.}
\end{figure}
%%%%%%%%%%%%%%%%% Fig .5 %%%%%%%%%%%%%%%%%%%%%%%%%%%
The same graph shows also the variances of these dynamical quantities. 
Assuming that the fragment consists of nucleons which are initially
in each direction distributed according to a Gaussian distribution $$f(p) = 
({1\over 2\pi\sigma^2})^{1/2}
e^{-p^2/2\sigma^2}$$ with $\sigma^2 = {<p^2>\over 3} = {p_{Fermi}^2\over 5} = 
12.5\times 10^{-3} \ (GeV/c)^2$ we expect $\sigma(y) = .05$ and 
$\sigma(p_x) = .0469\ GeV/c$. Initially we indeed observe these values
for both the transverse momentum and the rapidity and whatever the class
considered. Moreover, the large width of the ``all nucleons'' rapidity 
distribution of Fig. 5
clearly results from a sum of three distributions, each having
a different $<y>$, but the same width $\sigma(y)$.

Once the reaction occurs, 
the interaction between projectile and target does
not only change the mean value but also increases the variance
considerably. In other words, this interaction does not affect all nucleons  
in the same way. Thus we can conclude that the
interaction between target and projectile widens the momentum distribution 
also of those nucleons which are finally entrained in a projectile-like IMF. 

In the past the variance of the observed IMF momentum distribution 
has sometimes been considered as a consequence of an instantaneous breakup of the system.
As calculated by  Goldhaber \cite{gol} the instantaneous break off of 
fragments of size A from a system of size M yields 
a Gaussian momentum distribution of these
fragments with a mean square momentum of 
\be
<p^2> = {3k_{Fermi}^2\over 5} A  { M-A\over  M-1}.
\ee
The above discussion has shown that the underlying physics is more complicated
than considered in the simple Goldhaber model. Before being part of a fragment
the nucleons change their momenta due to the interaction between projectile
and target. As we have discussed this interaction leads to an increase of
the width of the momentum distribution. Therefore the widths of the
momentum distribution found in experiments are systematically larger than
those predicted by the Goldhaber formula. A fit of eq.~5
to the ALADIN data \cite{schu} yields a Fermi momentum of $k_{Fermi} = 370 MeV/c$ which
is about 50 \% larger than that extracted from electron scattering. 

Despite that quantitative discrepancy the physics behind the Goldhaber 
model corresponds much more to the result we obtain than the assumption that
the system approaches thermal equilibrium. Like in the Goldhaber model
we find that the variance of the momentum distribution changes little
in the course of the interaction, whereas thermodynamics approaches assume
that is it generated by randomizing collective motion. Like in the
Goldhaber model we observe that the internal excitation energy of the
fragment is almost independent of its center of mass motion,
whereas in thermodynamics both are connected by a common temperature.
And last, but not least, we see that an important part of the variance of
the momentum distribution is due to the initial Fermi motion in a cold
nucleus at zero temperature whereas the thermodynamical models assume
that at zero temperature the nucleons have also zero velocity. 

The last row of fig.~5 displays on the left hand side the average total
energy/nucleon $E_{int}/N$ of the cluster. It is defined in eq.~3. For 
negative values $E_{int}/N$ corresponds to the binding energy per nucleon.
Initially the value of $E_{int}/N$ for those fragments consisting of 
projectile nucleons
only is - 1.55 MeV (as compared to -6.85 average binding energy of the
projectile nucleons) because the nucleons which finally belong to a given
fragment are not very close in coordinate space and therefore the potential
energy is not very high. If we include the fragments which contain 
target-like nucleons, this mean value increases to +5 MeV. When the violent phase of 
the reaction is over this value has decreased to
-3 MeV. This means that the target like nucleons which are finally entrained
in the projectile-like fragments have suffered collisions which have
placed them in momentum space close to the projectile nucleons with
whom they will form a fragment. At the moment the fragments are formed 
their binding energy is about 4.5 MeV/N lower than that of the ground state.
For matter of comparison with the experimental data
we apply the Fermi gas formula (although the
system is not in equilibrium) and obtain with $a=0.15 MeV^{-1}$ an 
apparent temperature of $"T" = \sqrt{E/a}= 5.5 MeV$. 
This is almost exactly the value one extracts
for the internal temperature of the IMF's from the experimental isotope ratios 
or from the experimentally observed  populations of 
particle unstable states as compared to the ground state.

The right hand side of the last row displays
\begin{equation} 
<E_{kin}/N> =  <\frac{1}{N^f}\left[\sum_{i=1}^{N^f} 
\frac{(\vec p_i - \vec p_{cm})^
{2}}{2m}\right]>
\end{equation}
the average kinetic energy of the nucleons belonging to a given cluster.
We display the average over all fragments. 
This quantity is a mixture of the Fermi energy and the system's excitation energy. 
We see that this quantity changes little and is close to the value one
obtains from the Fermi motion $(<E_{Kin}/N> \approx 20 MeV)$. Consequently,
the change of the average energy $E_{int}/N$ displayed on the left hand side
is mainly due the fact that the nucleons, which will finally form a fragment, 
come closer together and hence the potential energy increases. 

In conclusion we have shown that at energies around 1~GeV/N IMF's are 
dominantly formed in semiperipheral reactions. Before the 
formation the interaction between the projectile and target has 
considerably changed the momentum distribution of those nucleons which
are finally part of projectile-like IMF's. The width of the momentum
distribution has increased and the average momentum has decreased.

Nucleons which come close in coordinate {\it and} momentum space
during this first interaction phase between projectile
and target may form a fragment. This condition implies that the 
internal excitation
energy remains small. Inversely, nucleons with large relative momenta 
do not form fragments even if they are close together in coordinate space. 
Hence
{\it all observables, which are sensitive to the internal excitation
energies should display an apparent temperature of about 5 MeV}.    

The almost linear dependence of the variance of the center of mass motion 
of the fragments on the mass shows that the fragment momentum distribution is 
a convolution of the momentum distribution of the nucleons 
at the time of fragment formation. Due to the interaction
of projectile and target and due to the fact that the average momentum of the
fragments depends on the number of target nucleons entrained the width of this
distribution
at the time of fragment production is larger than initially. Therefore
the Goldhaber formula underestimates the mean square momentum of the fragments.
Because also thermodynamics predicts a linear dependence of the variance
of the momentum distribution on the fragment mass this dependence may be also
interpreted in a model which assumes thermal equilibrium of the system. 
From the mass dependence of the slope one obtains, under this assumption,
an apparent temperature of 30 MeV \cite{schu}. 

The QMD simulations, with the conjecture that the true fragment distribution
is obtained after 60 fm/c,  reproduce quantitatively both observed phenomena,
those which are usually connected with a low temperature (multiplicity
distribution and internal excitation) as well as those which display
a much larger apparent temperature like the kinetic energy distribution 
of projectile like fragments. This is possible because the system does
not pass through a state of thermal equilibrium despite the fact that many 
observables show a form expected for a system in thermal equilibrium.
Whether it reproduces the observed energy independence of the fragment
multiplicity is presently under investigation

%%%%%%%%%%%%%%%%% Fig .6 %%%%%%%%%%%%%%%%%%%%%%%%%%%
\begin{figure}[htb]
\centerline{\psfig{figure=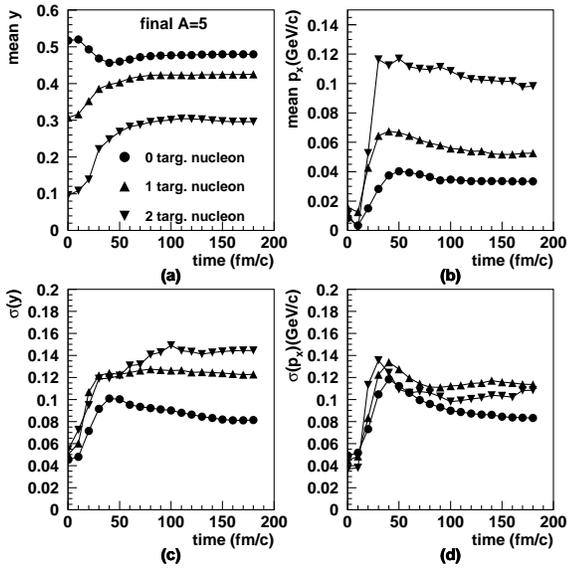,width=0.95\hsize}}
\caption{Evolution of the collision of Au-Au at 600 MeV/N and at
an impact parameter of b = 8 fm. We display the mean values and 
the variances of the rapidity and the in-plane transverse momentum 
distributions for three different classes of A=5 projectile like fragments.}
\end{figure}
%%%%%%%%%%%%%%%%% Fig .6 %%%%%%%%%%%%%%%%%%%%%%%%%%%

{\it Acknowledgment}: We would like to thank Drs. M\"uller, Sch\"uttauf
and Trautmann for many discussions and for providing us with experimental
data of the ALADIN collaboration.

%%\vspace*{0.2cm}\\

%%%%%%%%%%%%%%%%%%%%%%%% newpage 

%%%%%%%%%%%%%%%%%%%%%%%%%%%%%

\end{document}